# Critical behavior and magnetocaloric effect in Tsai-type 2/1 and 1/1 quasicrystal approximants

Farid Labib[1*], Takafumi D. Yamamoto[2], Asuka Ishikawa[1] and Ryuji Tamura[2]

[1] *Research Institute of Science and Technology, Tokyo University of Science, Tokyo 125-8585, Japan*
[2] *Department of Material Science and Technology, Tokyo University of Science, Tokyo 125-8585, Japan*

Stable Tsai-type quinary 1/1 and 2/1 approximant crystals (ACs) with chemical compositions $Au_{56.25}Al_{10}Cu_7In_{13}Tb_{13.75}$ and $Au_{55.5}Al_{10}Cu_7In_{13}Tb_{14.5}$, respectively, exhibiting ferromagnetic (FM) long-range orders were successfully synthesized and studied for their magnetic properties and magnetocaloric effect. The 1/1 and 2/1 ACs primarily differ in their long-range atomic arrangement and rare earth (RE) distribution, with the latter approaching quasiperiodic order while still preserving periodicity. Analyses based on the scaling principle and Kouvel-Fisher (KF) relations suggested mean-field-like behavior near Curie temperatures in both compounds. From magnetization measurements and Maxwell's equation, a magnetic entropy change ($\Delta S_M$) of −4.3 and −4.1 J/K mol Tb were derived under $\mu_0 \Delta H = 7$ T for the 1/1 and 2/1 ACs, respectively. The results indicated prominent role of intra-cluster magnetic interactions on critical behavior and $\Delta S_M$ of the Tsai-type compounds.

[*]Corresponding author:
labib.farid@rs.tus.ac.jp

## I. INTRODUCTION

Quasicrystals (QCs) are aperiodically ordered compounds that generate non-periodic Bragg reflections with 5-fold rotational symmetry in their diffraction patterns [1,2]. They are amongst the latest discoveries in the field of condensed matter physics that have attained considerable attention. Approximant crystals (AC), on the other hand, are closely related phases to QCs sharing the same rhombic triacontahedron (RTH)-shaped multi-shell polyhedron, also known as Tsai-type cluster (depicted in Fig. 1a), arranged periodically in the physical space. The lowest order AC, i.e., 1/1 AC, typically crystallizes in a cubic lattice and a space group $Im\bar{3}$ [3], even though other lattice symmetries may appear depending on the order/disorder of the innermost unit of the RTH cluster [4–8]. Their lattice parameter increase with the order of AC; the higher their order (following the rational approximation of golden mean $\tau = 1.61803$), the more their atomic structure resembles to that of icosahedron quasicrystals ($i$QCs) and the larger their lattice parameter become [9]. In the course of structure evolution from 1/1 AC to $i$QC, the RTH clusters are re-arranged until a long-range five-fold symmetry is obtained in the $i$QC (as schematically shown in Fig. 1b). The intriguing fact about the higher order ACs (starting from 2/1 AC and above) is that they share all the necessary building units to construct the $i$QCs, i.e., RTH cluster, obtuse rhombohedra (OR) and acute rhombohedra (AR), as depicted on the right-hand side of Fig. 1a. In particular, the latter unit that comprises two additional rare earth (RE) elements along its long body diagonal axis may alter the magnetic interactions by changing the distance distribution between REs via Ruderman-Kittel-Kasuya-Yosida (RKKY) mechanism. In this sense, the 2/1 AC serves as an ideal choice in studying the effect of structural aperiodicity on magnetism which is one of the hot topics in condensed matter physics.

When it comes to magnetism in Tsai-type alloys, several milestones should be mentioned starting from the first discovery of long-range magnetic order in $Cd_6Tb$ 1/1 AC more than a decade ago [10]. Later, significant advancements (both theoretically and experimentally) have been made unveiling the physics of such phenomenon, among which the establishment of the electron concentration (expressed as electron-per-atom $e/a$) dependence of the magnetic ground states [11] and the discoveries of exotic noncoplanar ferromagnetic (FM) [12] and antiferromagnetic (AFM) orders [13] in Tsai-type ACs could be mentioned. Following a number of reports about long-range magnetic order establishments in ACs [11,14–17], the most recent discoveries of long-range FM order in Au-Ga-*RE* (RE = Gd, Tb) [18] and Au-Ga-Dy [19] $i$QCs finally resolved a long-standing debate whether long-range order can survive in quasiperiodic environment.

These above discoveries have now made it possible to investigate the fundamental physics of these long-range orders, especially the critical behavior near the FM transition temperature in Tsai-type $i$QCs and ACs. Furthermore, the study of the magnetocaloric effect (MCE), which is a phenomenon directly related to the critical behavior, has become quite an emerging and underexplored research topic in Tsai-type $i$QCs and ACs. Non-coplanar spin configurations observed in 1/1 ACs [13,20–22] may hold potential for significant MCE effects yet to be discovered. In ref. [23], magnetic entropy change ($\Delta S_M$) in $Au_{64}Al_{22}RE_{14}$ (RE = Gd, Tb, and Dy) 1/1 ACs was studied and approximately 30% improvement was noticed in $\Delta S_M$ for Gd-contained 1/1 AC compared to Tb- and Dy-contained ones. From this observation, a favorable effect of a weak single-ion anisotropy in a $Gd^{3+}$ on MCE behavior in Tsai-type compounds was inferred.

The present work aims to investigate the impact of structural evolution towards aperiodicity on the critical behavior and $\Delta S_M$ of the ferromagnetically ordered Tsai-type compounds. Here, we report successful synthesis of 1/1 and 2/1 ACs with FM ground state and

…

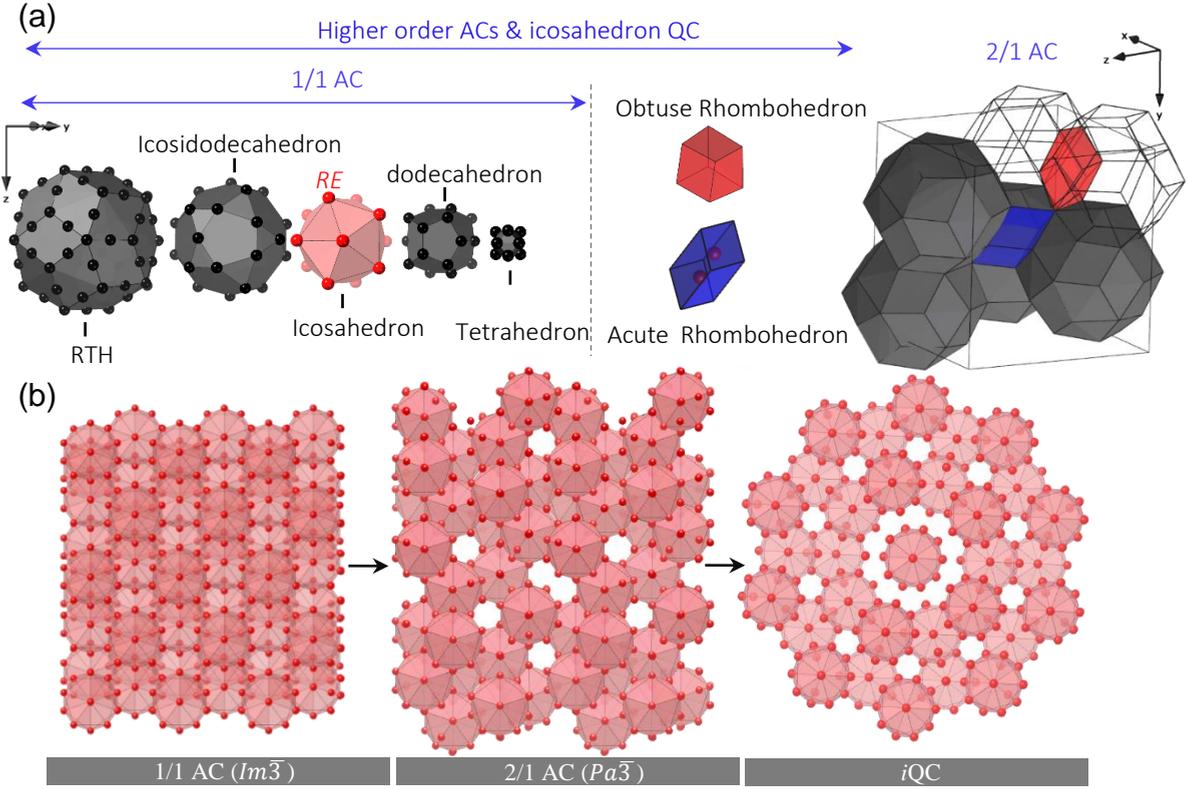

FIG. 1. (a) (left) A typical shell structure of the Tsai-type $i$QC. From the outermost shell to the center: a rhombic triacontahedron (RTH) with 92 atoms sites, an icosidodecahedron (30 atomic sites), an icosahedron (12 atomic sites), a dodecahedron (20 atomic sites) and an innermost tetrahedron (4 atomic sites). (right) A typical arrangement of RTH clusters within a unit cell of 2/1 AC including an AR unit that fills the gap between RTH clusters and an OR as a shared unit of the RTH clusters along three-fold axes. (b) Configuration of the RE elements within the structure of the 1/1 AC, 2/1 AC and an $i$QC looking along five-fold axis of $i$QC (or pseudo-file-fold axes of ACs). The depth of the cut segment in physical space in (b) is 46 – 48 Å.

comparable $e/a$ parameters by engineering the structure of the ternary Au-Al-Tb 1/1 AC mother alloy via expansion of the synthesis space to quinary Au-Cu-Al-In-Tb system. Our strategy involved a comparative analysis of the magnetic properties, critical behavior and $\Delta S_M$ in the synthesized 1/1 and 2/1 ACs. If differences between the two detected, they could be ascribed to the structural aperiodicity. Otherwise, it could be attributed to the shared building unit common to both ACs, i.e., RTH cluster.

## II. EXPERIMENT

To prepare polycrystalline quinary 1/1 and 2/1 ACs, appropriate amounts of mono-valent Cu and tri-valent In (iso-valent to Au and Al, respectively) were added to the ternary Au-Al-Tb 1/1 AC mother alloy, resulting in compounds with nominal compositions of $Au_{56.25}Al_{10}Cu_7In_{13}Tb_{13.75}$ and $Au_{55.5}Al_{10}Cu_7In_{13}Tb_{14.5}$, respectively. Nearly 0.7 at.% higher concentration of Tb in the 2/1 AC is due to the existence of AR unit that incorporates additional RE elements along the long body-diagonal axis (see Fig. 1). The arc-melting technique followed by 100 hours of isothermal annealing at 773 K and 673 K (for 1/1 and 2/1 ACs, respectively) was employed for synthesizing the alloys. For phase identification, powder X-ray diffraction (XRD) was employed using Rigaku SmartLab SE X-ray Diffractometer with Cu-K$\alpha$ radiation. For phase characterization, selected area electron diffraction (SAED) patterns were obtained using a transmission electron microscopy (TEM) instrument JEM-2010F, located at the Advanced Research Infrastructure for Materials and Nanotechnology (ARIM) at the University of Tokyo. Superconducting quantum interference device (SQUID) magnetometer (Quantum Design, MPMS3) was utilized to examine the bulk magnetization under zero-field-cooled (ZFC) and field-cooled (FC) conditions within a temperature range of 1.8 K to 300 K and in external dc fields up to $7\times10^4$ Oe. Additionally, specific heat measurements were conducted in a temperature range of 2K − 50 K by a thermal relaxation method using a Quantum Design PPMS. The magnetic entropy change ($\Delta S_M$) was determined by analyzing a series of temperature-dependent magnetization ($M$ vs. $T$) curves under various magnetic fields within the temperature range of 2 K to 120 K, employing a thermodynamic Maxwell relation.

## III. RESULTS

Figure 2 displays powder XRD patterns of the synthesized 1/1 and 2/1 ACs, along with their Le Bail fittings. The fittings were conducted using the Jana 2006 software suite [24], assuming the space group $Im\bar{3}$ and $Pn\bar{3}$, respectively. The red, black, and blue lines represent the observed ($I_{obs}$), calculated ($I_{cal}$) peak intensities, and the difference between the two, respectively. The expected Bragg peak positions are indicated by the green bars. From the fittings, the lattice parameters of the 2/1 and 1/1 ACs were

…

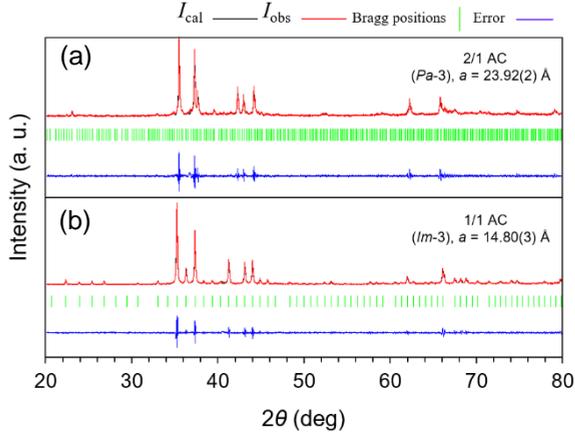

FIG. 2. Le Bail fitting of the powder XRD patterns of (a) $Au_{55.5}Al_{10}Cu_7In_{13}Tb_{14.5}$ 2/1 AC and (b) $Au_{56.25}Al_{10}Cu_7In_{13}Tb_{13.75}$ 1/1 AC. The observed ($I_{obs}$), calculated ($I_{cal}$) peak intensities, the difference between the two and the expected Bragg peak positions are represented by red, black, blue lines and green bars, respectively.

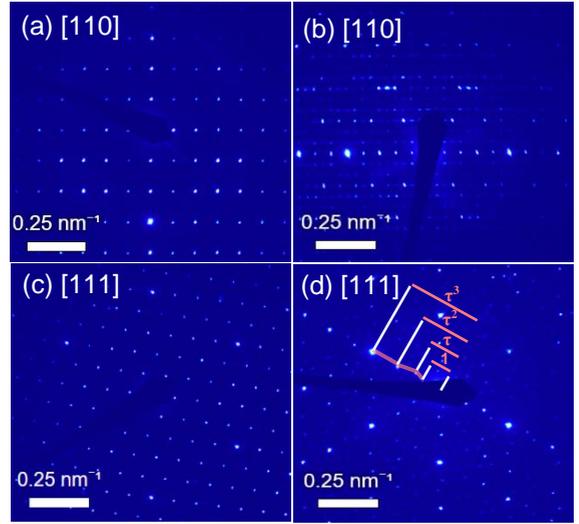

FIG. 3. SAED patterns obtained from (a,c) 1/1 AC and (b,d) 2/1 AC along incident axes perpendicular to (a,b) [110] and (c,d) [111] directions.

determined to be 23.92(2) and 14.80(3) Å, respectively. The excellent agreement between the calculated and experimental peaks in Fig. 2 confirms high purity of the synthesized samples, making them suitable for further analysis of their physical properties.

Figure 3 showcases the SAED patterns obtained from incidence axes perpendicular to (a, b) [110] and (c, d) [111] directions of the (a, c) 1/1 AC and (b, d) 2/1 AC. The patterns reveal the cubic lattices and the space groups $Im\bar{3}$ and $Pn\bar{3}$ for the 1/1 and 2/1 ACs, respectively, validating the Lebail fittings in Fig. 2 where these conditions are applied. Furthermore, in the SAED pattern taken along the [111] axis of the 2/1 AC in Fig. 3d, a clear zig-zag-like arrangement of the diffraction spots being deviated from a perfect τ scaling is observed, indicating the presence of a linear phason strain within the structure, which is a characteristic feature of the 2/1 AC. These structural features enable us to investigate the effect of structural evolution towards quasiperiodicity on the physical properties and MCE behavior of these compounds.

In Fig. S1, the high-temperature inverse magnetic susceptibility ($H/M$) of the compounds is shown within a temperature range of 1.8–300 K. The results demonstrate a linear behavior in both ACs, fitting well to the Curie–Weiss law: $\chi(T) = N_A \mu_{eff}^2 \mu_B^2 / 3k_B(T - \theta_w) + \chi_0$ where $N_A$, $\mu_{eff}$, $\mu_B$, $k_B$, $\theta_w$, and $\chi_0$ denote the Avogadro number, effective magnetic moment, Bohr magneton, Boltzmann constant, Curie-Weiss temperature, and the temperature-independent magnetic susceptibility, respectively. By extrapolating a linear least-squares fitting within a temperature range of 100 K < $T$ < 300 K, the estimated $\theta_w$ values of +17.8(5) and +14.6(5) K are derived for the 1/1 and 2/1 ACs, respectively, with $\chi_0$ being nearly zero. The obtained $\mu_{eff}$ values in both compounds are within a range of 9.57 – 9.71 $\mu_B$, close to 9.72 $\mu_B$, i.e., the calculated value for free $Tb^{3+}$ ions defined as $g_J(J(J+1))^{0.5}$ $\mu_B$ [25], indicating the localization of magnetic moments on $Tb^{3+}$ ions. The temperature dependence of the dc magnetic susceptibility ($M/H$) of the 1/1 and 2/1 ACs is illustrated in Fig. 4 within the temperature range of 1.8 – 30 K under FC (filled circles) and ZFC (unfilled circles) modes. The inset provides a first derivative of $M/H$ versus temperature. Both 1/1 and 2/1 ACs exhibit a relatively sharp rise in their magnetic susceptibility below $T_C$ = 15.9 K and 14.2 K, respectively, which is attributed to the pinning of magnetic domain walls during the FC process: a common phenomenon in ferromagnets.

Figure 5 depicts magnetization curves ($M$-$H$) measured at 1.8 K for the 1/1 (red) and 2/1 (blue) ACs. The inset displays complete $M$-$H$ loops in a magnetic field range of −0.4 to 0.4 T. The main panel of Fig. 5 demonstrates a rise of magnetization with increasing magnetic field in both ACs, reaching approximately 6 $\mu_B$/$Tb^{3+}$ (about 75% of the total moment of a free $Tb^{3+}$ ion) at 7T. The unsaturation to a full moment of $Tb^{3+}$ at high magnetic fields is a common feature among ferromagnetically ordered non-Heisenbeg ACs and is attributed to a strong uniaxial anisotropy of the spins with non-zero orbital angular momentum, as described elsewhere [26,27]. Interestingly, the 1/1 AC exhibits a relatively larger remanence magnetization and coercivity compared to the 2/1 AC, as seen in the inset of Fig. 5. Although there is no report about the magnetic structure of the higher order ACs or $i$QCs to date, under the assumption of identical spin arrangement on the icosahedron vertices in the lower and higher order FM ACs, i.e., non-coplanar whirling structure formerly reported in the Au-Si-$RE$ (RE = Tb, Ho) [20,21], the observed minute difference in the remanence magnetization and coercivity may be correlated to the presence of additional RE elements inside the AR unit in the atomic structure of higher order ACs and *i*QCs (see Fig. 1). These additional RE elements could create competing interactions between magnetic moments or increase randomness between FM and AFM interactions through RKKY indirect coupling. Other possibility could be the differences in the chemical disorder of the present 1/1 and 2/1 ACs induced by their slight compositional dissimilarities, which may alter the magnetic moments through the CEF effect, as discussed elsewhere [28], and lead to a difference in remanence magnetization and coercivity between the two ACs.

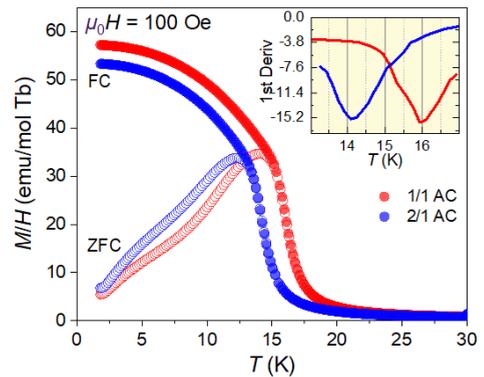

FIG. 4. Low temperature magnetic susceptibility $M/H$ of the 1/1 AC (represented by red) and 2/1 AC (represented by blue) under field-cooled (FC) and zero-field-cooled (ZFC) modes.

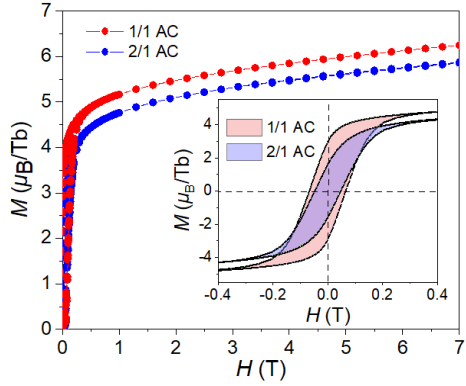

FIG. 5. Field dependence magnetizations of 1/1 AC (red) and 2/1 AC (blue) measured at 1.8 K up to $H_{dc}$ = 7 T. In the inset, $M$-$H$ loops of both ACs from −0.4 T to 0.4 T is provided.

In Fig. 6, the temperature dependence of the specific heat divided by temperature ($C_p/T$) for the 1/1 AC (a red line) and 2/1 AC (a blue line), as well as the non-magnetic $Au_{56}Al_{10}Cu_7In_{13}Y_{14}$ 2/1 AC (shown in green) is provided within a temperature range of 1.8−50 K. The positions of pronounced anomalies at $T$ = 15.8 K and 14.3 K are in perfect agreement with the $T_C$ values estimated from the magnetization data, confirming the establishment of long-range orders in the studied ACs. In the paramagnetic region (above $T_C$), the $C_p/T$ curves of the ACs overlap with each other, indicating nearly identical lattice contributions to $C_p$ in lower and higher order ACs despite differences in their long-range atomic order. The inset of Fig. 6 displays the magnetic entropy of the ACs with the error bars stemming from uncertainty in the chemical composition and/or the measured weight of the samples. The magnetic entropy in the samples has been estimated from the magnetic contribution to heat capacity ($C_M$), derived by subtracting the nonmagnetic ($C_{NM}$) contribution using the data of the $Au_{56}Al_{10}Cu_7In_{13}Y_{14}$ 2/1 AC and the following relation:

$$S_M = \int_0^T C_M/T \; dT \quad (1)$$

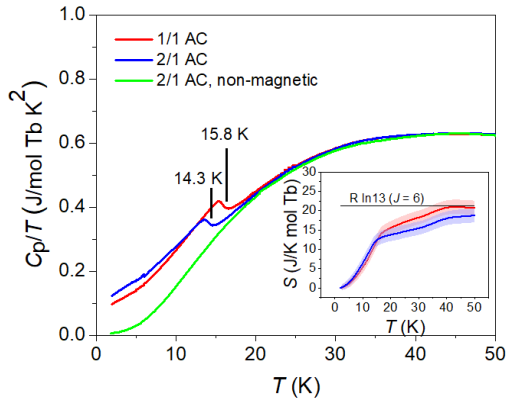

FIG. 6. The temperature dependence of the specific heat for 1/1 AC (represented by red), 2/1 AC (blue) and non-magnetic $Au_{56}Al_{10}Cu_7In_{13}Y_{14}$ 2/1 AC (green) divided by temperature $C_p/T$. The inset shows the magnetic entropy with the shaded area representing the error bar.

As seen in the inset of Fig. 6, a zero-field magnetic entropy $S_M$ approaches $R \ln(2J+1) = 13$ ($J$: total angular momentum = 6 for $Tb^{3+}$) in both ACs, which is a maximum magnetic entropy expected for per mole of $Tb^{3+}$ with $R$ being a universal gas constant.

Moving forward, we examine the critical behavior of the present FM 1/1 and 2/1 ACs. Figure S2 illustrates the $H/M$ dependence of $M^2$ in the form of a standard Arrott plot [29]. The absence of a negative slope and/or an inflection point in Fig. S2 suggests the existence of second-order phase transitions in the 1/1 and 2/1 ACs. Based on the scaling principle, near the critical temperature ($T_C$) in the second-order phase transition, the following equations should hold [30]:

$$M_s(T) = M_0(-\epsilon)^\beta; \; \epsilon < 0; \; T < T_c \quad (2)$$

$$(H/M)_0(T) = (h_0/M_0)\epsilon^\gamma; \; \epsilon > 0; \; T > T_c \quad (3)$$

where $M_0$ and $h_0$ are critical amplitudes and $\epsilon$ is a reduced temperature $(T - T_C)/T_C$. The critical exponents $\beta$ and $\gamma$ correspond to the spontaneous magnetization $M_s(T)$ below $T_C$ ($H = 0$) and an initial inverse magnetic susceptibility $(H/M)_0(T)$ above $T_C$, respectively. Applying equations (2) and (3) and after a few cycles of iteration (refer to supplementary material for details), convergence in the optimal $\beta$ and $\gamma$ values is reached, reflecting the intrinsic behavior of the material. Figure S3 in the supplementary material co-plots the fitting results of the $M_s(T)$ and $(H/M)_0$ values to equations (2) and (3) in the final cycle, from which $\beta = 0.41(2)$; $\gamma = 0.89(3)$; $T_C = 16.14(4)$ K for the 1/1 AC and $\beta = 0.39(2)$; $\gamma = 1.09(3)$; $T_C = 14.32(7)$ K for the 2/1 AC are derived. In addition, the critical exponents near $T_C$ are determined using the Kouvel-Fisher (KF) equations [31] (refer to Fig. S4 in the supplementary material for details), from which $\beta = 0.39(1)$; $\gamma = 0.90(3)$; $T_C = 15.95(6)$ K for the 1/1 AC and $\beta = 0.39(2)$; $\gamma = 1.06(4)$; $T_C = 14.39(5)$ K for the 2/1 AC are derived. The consistency of the critical exponents derived from KF analysis and equations (2) and (3) using the scaling principle indicates credibility of the performed analyses.

Figure 7 plots $M^{1/\beta} - (H/M)^{1/\gamma}$ for the 1/1 and 2/1 ACs near $T_C$ using the ($\beta;\gamma$) of (0.40, 0.90) and (0.39, 1.08) for the 1/1 and 2/1 ACs, respectively. Here, the average of the values obtained from the above analyses are considered for the $\beta$ and $\gamma$. Nearly parallel lines of the isotherms within a magnetic field range of 0.4 – 1.5 T (represented by colored segments in Fig. 7) with that corresponding to $T_C$ passing close to origin further confirm the credibility of the analyses and the derived critical exponents. The calculated $T_C$ for the 1/1 and 2/1 ACs falls within the range of 15.95 – 16.14 K and 14.32 – 14.39 K, respectively, being in accordance with $T_C$ estimated from magnetization and specific heat data. The insets in Fig. 7 plot $M/\varepsilon^\beta$ vs. $H/\varepsilon^{(\beta+\gamma)}$ on a logarithmic scale based on static-scaling hypothesis ($\frac{M(t,H)}{\epsilon^\beta} = f_\pm[H/\varepsilon^{(\beta+\gamma)}]$) whereby the collapsing

...

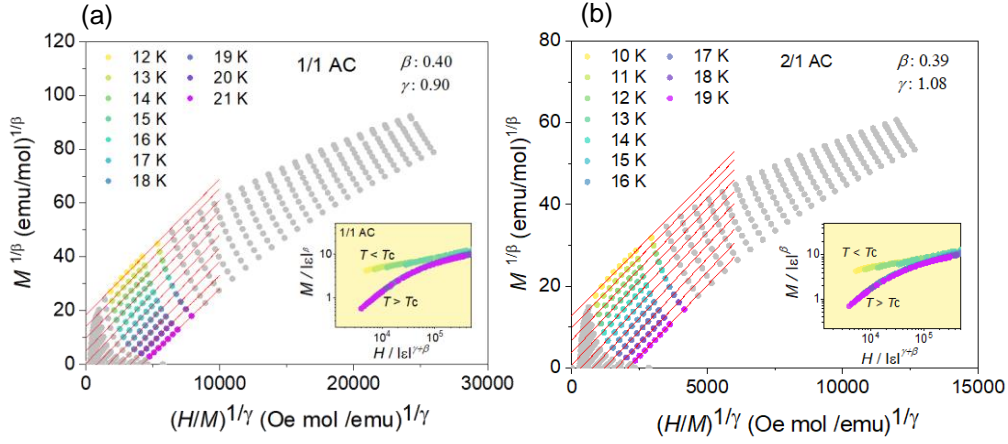

FIG. 7. The isotherms of $M^{1/\beta}$ vs. $H/M^{1/\gamma}$ for (a) 1/1 AC, (b) 2/1 AC. Nearly parallel linear behavior can be observed within the coloured segments of the isotherms corresponding to magnetic fields of $0.4\ \mathrm{T} < H < 1.3\ \mathrm{T}$. The insets show logarithmic scaling plot of $M/|\varepsilon|^{\beta}$ vs. $H/|\varepsilon|^{\beta+\gamma}$ in the critical region. The experimental points fall on two branches of a universal curve by using the derived values of the critical exponents.

of the data points on two branches, i.e., $f_+$ for $T > T_c$ and on $f_-$ for $T < T_C$ indicates adaptation of correct critical exponents [32].

Using the above analyses, the estimated critical exponents yield $n = \gamma/\beta = 2.16 - 2.33$ for the present ACs which are lower than those expected for the three-dimensional (3D) universality classes: $n = 3.80$ [29] (3D Heisenberg), $n = 3.82$ [29] (3D Ising), and $n = 4.00$ [33] (tricritical mean-field) but are close to the $n = 2.00$ predicted by the Landau mean-field model [29] indicating a mean-field-like nature of the phase transition near $T_C$. Note that the estimated critical exponents in the present ACs accord well with the $(\beta,\gamma) = (0.54, 0.89)$ [19] and $(0.47, 1.12)$ [34] obtained from bulk magnetization data in Au-Ga-Dy $i$QC and Au-Si-Gd 1/1 AC, respectively, and $\beta = 0.44(2)$ [22] and $\beta = 0.56(4)$ [35] estimated from neutron diffraction experiment in the $Au_{70}Al_{16}Tb_{14}$ and $Au_{65}Ga_{21}Tb_{14}$ 1/1 ACs, respectively.

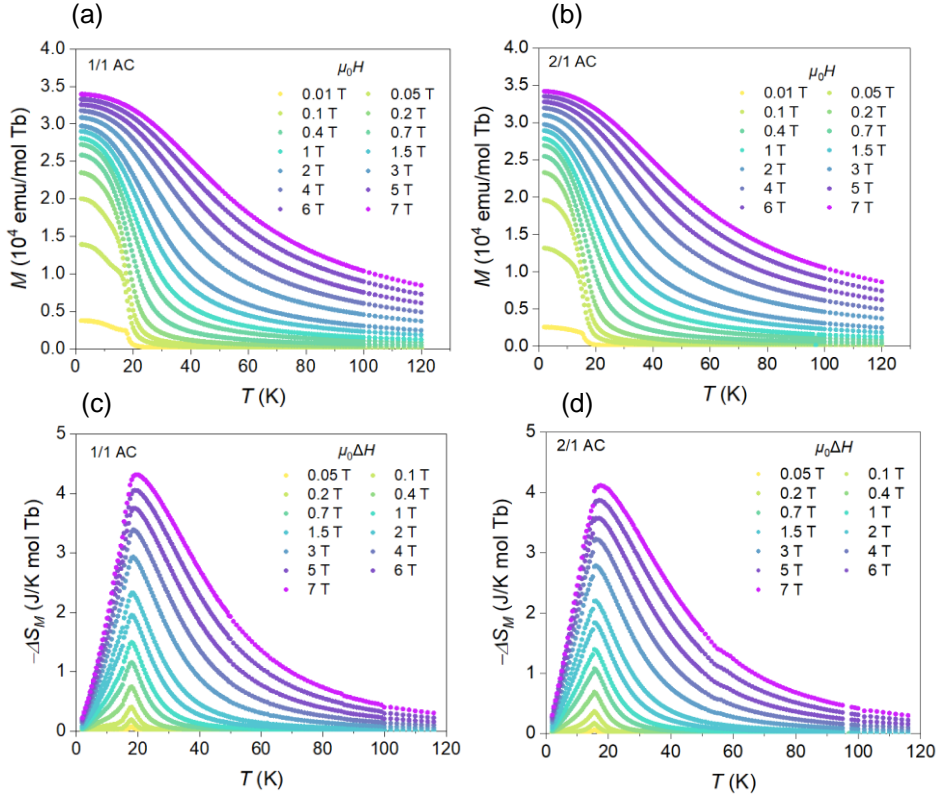

FIG. 8. Series of temperature dependence of magnetization under FC mode for (a) 1/1 AC, (c) 2/1 AC under magnetic field spanning from $0.01 - 7$ T. The corresponding $\Delta S_M$ are shown in (b) for 1/1 AC and (d) for 2/1 AC.

…

Table I. Comparison of the $T_C$, $-\Delta S_M$, and RCP under a field change of 5 T in the present ACs and previous reports.

| AC Type | Composition | Polycrystal / single crystal | $T_C$ (K) | $-\Delta S_M$ (J/K mol Tb) | RCP (J/mol Tb) | Reference |
|---|---|---|---|---|---|---|
| 1/1 | $Au_{56.25}Al_{10}Cu_7In_{13}Tb_{13.75}$ | Polycrystal | 15.9 | 3.8 | 115 | This work |
| 2/1 | $Au_{55.5}Al_{10}Cu_7In_{13}Tb_{14.5}$ | Polycrystal | 14.3 | 3.7 | 107 | This work |
| 1/1 | $Au_{66}Al_{20}Gd_{14}$ | Polycrystal | 27 | 5.3 | 135 | [23] |
| 1/1 | $Au_{66}Al_{20}Tb_{14}$ | Polycrystal | 15 | 3.9 | 105 | [23] |
| 1/1 | $Au_{66}Al_{20}Dy_{14}$ | Polycrystal | 9.5 | 4.3 | 107 | [23] |
| 1/1 | Au-Si-Gd | Single crystal | 16.9 | 5.4 | 130 | [34] |

In what follows, we investigate magnetic entropy of the present ACs. Figure 8 provides series of temperature dependence of the FC magnetization curves for (a) 1/1 AC and (b) 2/1 AC within a temperature range of 1.8 – 120 K and magnetic fields spanning from 0.01 – 7 T. The $-\Delta S_M$ variation with temperature for the 1/1 and 2/1 ACs (presented in Figs. 8c and d, respectively) is estimated using the thermodynamic Maxwell relation [36]:

$$\Delta S_M(T,H) = \mu_0 \int_{H_1}^{H_2} \left(\frac{\partial M(T,H)}{\partial T}\right)_H dH \quad (4)$$

with $M$ and $H$ representing the magnetization and the external magnetic field, respectively. Clearly, $-\Delta S_M$ in both ACs shows a maximum around $T_C$. Under $\mu_0 \Delta H = 7$ T, the $\Delta S_M$ of the 1/1 and 2/1 ACs amount to $-4.3$ and $-4.1$ J/K mol Tb, respectively. From Figs. 8c and d, relative cooling power (RCP), a measure of heat transfer between the hot and cold reservoirs, can be estimated using the following equation [36]:

$$RCP = -\Delta S_M \times \delta T_{FWHM} \quad (5)$$

where $\Delta S_M$ and $\delta T_{FWHM}$ correspond to maximum magnetic entropy change at given magnetic field and the full width at half maximum of the corresponding $\Delta S_M(T)$ curve, respectively. The RCP of the present 1/1 and 2/1 ACs versus $\mu_0 \Delta H$ is provided in Fig. S5 in the supplementary material. Table I lists estimated $T_C$, $-\Delta S_M$ and RCP values of the present 1ACs under $\mu_0 \Delta H = 5$ T together with those reported for the $Au_{64}Al_{22}RE_{14}$ (RE = Gd, Tb, Dy) [23] and Au-Si-Gd [34] 1/1 ACs for comparison. The results indicate comparable $\Delta S_M$ for the Tb- and Dy-contained ACs regardless of the alloy system being ternary or quinary. This may suggest an insignificant effect of chemical disorder, which essentially rises by increasing the number of the involved elements in the compound, on MCE behavior of Tsai-type compounds. In addition, no significant change is noticed between the $\Delta S_M$ of the higher and lower order ACs. Under the assumption of Au and Cu being mono-valent and In and Al being tri-valent [37], the $e/a$ of the present FM ACs ranges between 1.73 – 1.75 which is close to the $e/a = 1.72$ in the FM $Au_{64}Al_{22}RE_{14}$ 1/1 ACs [23]. This raises the possibility that the MCE in non-Heisenberg Tsai-type compounds with close $e/a$ parameters are comparable regardless of the order of AC.

In the subsequent discussion, we explore the factors contributing to the comparable magnetic properties of the present 1/1 and 2/1 ACs. In the Tsai-type systems, the exchange interaction between RE elements is primarily controlled by the RKKY mechanism, the strength of which is directly proportional to $f(x) =$

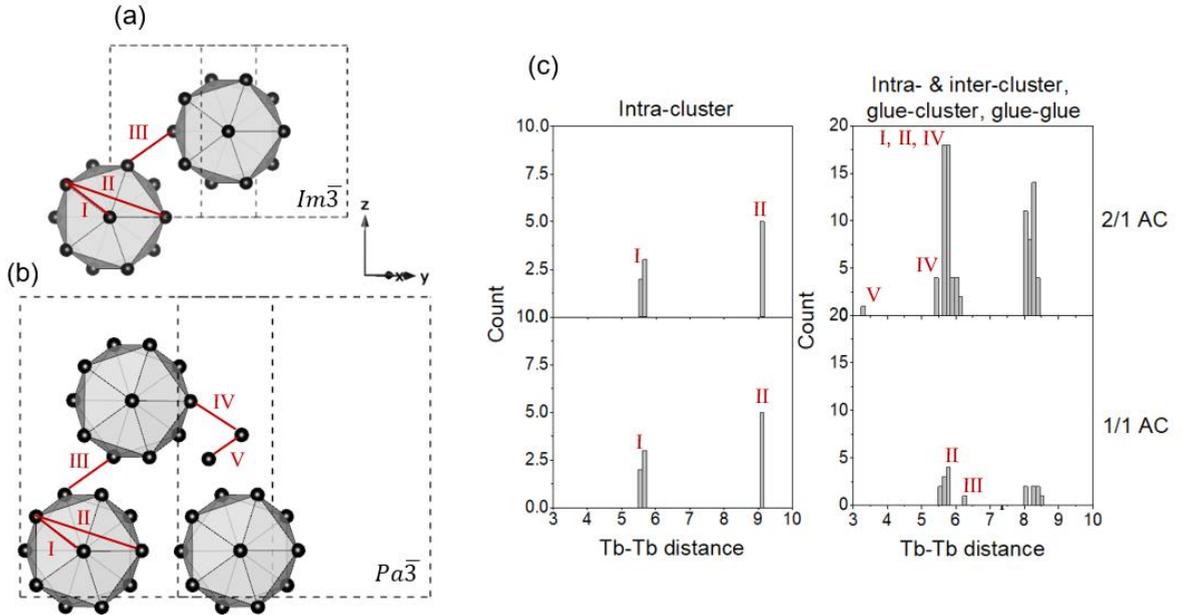

FIG. 9. Configuration of the RE elements on the icosahedron clusters in (a) 1/1 AC and (b) 2/1 AC. Distances (I) and (II) correspond to intra-cluster distances while (III), (IV) and (V) indicate inter-cluster, glue-cluster, and glue-glue distances, respectively. In (c) $r$ distribution in the 1/1 and 2/1 ACs are compared by considering 1) only intra-cluster distances and 2) all distances up to 10 Å covering nearest neighbor (NN) and next nearest neighbor (NNN). In (c) duplicate distances are removed and the bin size is set to 0.1186 Å.

...

($-x\cos x + \sin x$)/$x^4$ with $x = 2k_\text{F}r$ and $d$G. Here, the $k_\text{F}$, $r$ and $d$G correspond to Fermi wave vector, the distance between two spins, and de Gennes parameter (expressed as $(g-1)^2 J(J+1)$, with $g_J$ and $J$ denoting the Landé g-factor and the total angular momentum), respectively. In the present ACs, both being Tb-based, the $d$G parameter remains essentially identical. In addition, by applying the free electron approximation (based on which $k_\text{F} = (\frac{3\pi^2 N}{V})^{1/3}$ with $N/V$ being the number of free electrons per unit cell), the $k_\text{F}$ of the present 1/1 and 2/1 ACs are estimated to be $1.38 \times 10^{10}$ m$^{-1}$ and $1.40 \times 10^{10}$ m$^{-1}$, respectively, indicating their close proximity. The $r$ distribution, therefore, is a remaining parameter that could potentially affect the exchange interaction in the lower and higher order ACs.

Figures. 9a and b illustrate typical RE configurations on the icosahedron clusters in the 1/1 and 2/1 ACs, respectively. In the former, there exist 24 symmetrically equivalent RE sites in one unit cell, while in the latter, 104 RE sites with five distinct Wyckoff orbits are distributed. For simplicity, only a few icosahedrons are displayed in Figs. 9a and b. Figure 9c provides a comparison of the $r$ distribution in the 1/1 and 2/1 ACs considering 1) only intra-cluster distances (i.e., the distances within an icosahedron cluster labelled Ⅰ and Ⅱ in Fig. 9) and 2) all distances encompassing intra-cluster (Ⅰ and Ⅱ), inter-cluster (Ⅲ), glue-cluster (Ⅳ) and glue-glue (Ⅴ) types (see Fig. 9b). Evidently, intra-cluster distances remain identical in both ACs, whereas when all the distances are considered, the '$r$' distribution in the 2/1 AC becomes more complex. The proximity of the experimental results derived from the present 1/1 and 2/1 ACs, nevertheless, suggests that intra-cluster magnetic interactions may play a dominant role in determining magnetic properties and $\Delta S_\text{M}$ in Tsai-type compounds. This conclusion is, however, contingent upon the assumption of identical spin configurations on a single icosahedron cluster in the 1/1 and 2/1 ACs, which is yet to be confirmed experimentally.

## CONCLUSION

This study compares the critical behavior and magnetocaloric effect of Tsai-type 1/1 and 2/1 ACs in the quinary Au-Cu-Al-In-Tb system, using bulk dc magnetic susceptibility and specific heat measurements. The analyses based on the scaling principle and Kouvel-Fisher relations revealed mean-field-like behavior near $T_\text{C}$ for both ACs. Through the Maxwell relation and magnetization measurements, the $\Delta S_\text{M}$ of $-3.8$ and $-3.7$ J/K mol Tb and RCP of 115 and 107 J/mol Tb were revealed under $\mu_0 \Delta H = 5$ T for the 1/1 and 2/1 ACs, respectively, comparable to the $\Delta S_\text{M} = -3.9$ J/K mol Tb and RCP = 115 J/mol Tb reported in the ternary Au$_{66}$Al$_{20}$Tb$_{14}$ 1/1 AC [23]. The results suggest that neither structural evolution towards aperiodicity nor chemical disorder plays a noticeable role in $\Delta S_\text{M}$ of Tsai-type compounds. This indicates that $\Delta S_\text{M}$ in Tsai-type compounds primarily arises from intra-cluster magnetic interactions rather than inter-cluster or other types of interactions.

## ACKNOWLEDGMENT

The authors acknowledge Dr. Akiko. T. Saito in magnetic and spintronic materials research center, National Institute for Materials Science (NIMS), Japan, for the heat capacity measurements. This work was supported by Japan Society for the Promotion of Science through Grants-in-Aid for Scientific Research (Grants No. JP19H05817, JP19H05818, JP19H05819, JP21H01044, JP22H00101, JP22H04582) and Japan Science and Technology agency, CREST, Japan, through a grant No. JPMJCR22O3.

## RERERENCES